\documentclass[12pt,preprint]{aastex}
\usepackage{amssymb}
\usepackage{lscape}

\shorttitle{MODIFIED SYNCHROTRON MODEL FOR KNOTS OF M87}
\shortauthors{Liu and Shen}

\begin{document}


\title{A MODIFIED SYNCHROTRON MODEL FOR KNOTS IN THE M87 JET}


\author{Wen-Po ~Liu\altaffilmark{1,2}, Zhi-Qiang ~Shen\altaffilmark{1}}


\altaffiltext{1}{Shanghai Astronomical Observatory, Shanghai 200030,
P.R.China; wpliu@shao.ac.cn}
\altaffiltext{2}{The Graduate School of
Chinese Academy of Sciences, Beijing 100049, P.R.China}


\begin{abstract}
For explaining the broadband spectral shape of knots in the M87
jet from radio through optical to X-ray, we propose a modified
synchrotron model that considers the integrated effect of particle
injection from different acceleration sources in the thin
acceleration region. This results in two break frequencies at two
sides of which the spectral index of knots in the M87 jet changes.
We discuss the possible implications of these results for the
physical properties in the M87 jet. The observed flux of the knots
in the M87 jet from radio to X-ray can be satisfactorily explained
by the model, and the predicted spectra from ultraviolet to X-ray
could be further tested by future observations. The model implies
that the knots D, E, F, A, B, and C1 are unlikely to be the
candidate for the TeV emission recently detected in M87.
\end{abstract}



\section{Introduction}

The giant radio galaxy M87, situated nearly at the center of the
Virgo cluster at a distance of 16 Mpc (e.g. Macri et al. 1999),
has been studied extensively from the radio (e.g., Owen et al.
1989 and references therein; Biretta et al. 1995; Sparks et al.
1996), optical (e.g. Perlman et al. 2001a, hereafter P01; Biretta
et al. 1991), UV (Waters \& Zepf 2005, hearafter WZ05), to X-ray
(e.g., Marshall et al. 2002, hereafter M02; Wilson \& Yang 2002).
The predicted TeV gamma-ray (Bai \& Lee 2001) and the possible
position where the TeV emission from the M87 jet (Georganopoulos
et al. 2005; Cheung et al. 2007) have also been confirmed at a
high confidence level by the HESS observations (Aharonian et al.
2003 $\&$ 2006). In particular, Perlman \& Wilson (2005, hereafter
PW05) analyzed diagnostics and physical interpretation of the
X-ray emissions from the M87 jet in detail. Many other
observations have also been done, such as photometric surveys of
the jet, polarimetry maps of the jet (Perlman et al. 1999).

In the M87 jet, the Optical-X-ray spectral index $\alpha_{\rm ox}$
($S_{\nu}\propto \nu^{-\alpha}$) increases from $1.2-1.4$ at
$2''-7''$ from the nucleus (knots D and E) to $1.7-1.9$ at
$15''-18''$ from the nucleus (knots B and C1). The Chandra data
confirm steep X-ray spectra of the knots in the M87 jet
($\alpha_{\rm x}>\alpha_{\rm r}$) and are thus consistent with a
synchrotron origin for the X-ray jet emission (M02; Wilson $\&$
Yang 2002; PW05). So far, there have been three standard
theoretical synchrotron models. The KP model (Kardashev 1962,
hereafter K62; Pacholczyk 1970) assumes that the source of the
emission is a single burst of energetic electrons with an
isotropic pitch-angle distribution and thus no scatters. Because
of the likely scattering of relativistic particles by
hydromagnetic waves (e.g., Wentzel 1977), the KP model is
physically unreasonable, and therefore will not be mentioned
further in this paper. The JP model (Jaffe \& Perola 1973) assumes
the same initial conditions as those of the KP model, but allows
scattering in the pitch-angle distribution so that it can maintain
an isotropic distribution all the time. The resulting spectrum is
an essentially exponential rollover above the synchrotron loss
break frequency. The continuous injection (CI) models (K62;
Heavens \& Meisenheimer 1987, hereafter HM87) assume that a power
law distribution of relativistic particles is being continuously
added to the emitting region, but the CI model of HM87 further
takes the advective transport of the accelerated electrons
downstream into account. The CI model of K62 has the similar
spectral shape to the one of HM87.

But these synchrotron models also have some problems. As shown in
WZ05, considering the X-ray flux, the CI model is only applicable
for the inner knots (knots D and E), and explains the UV turnover
(Fig 1), but for other outer knots (such as knots F and A) this
model systematically over-predicts the observed UV turnover. And
PW05 also found that such a model cannot explain the spectral
index at X-ray. Without considering the X-ray data, the predicted
X-ray flux by the CI model is higher than the observed one, but
the exponential high-energy rollover of JP model underpredicts the
X-ray flux by many orders of magnitude and the slope at X-ray is
much larger than the observed. The two standard models can't fit
the X-ray flux and the spectral index at X-ray under the single
index of the electron energy spectrum at injection and single
emission process. We summarize the criteria for fitting the
spectral energy distributions (SEDs) of the knots in the M87 jet
discussed in WZ05 and PW05$\colon$ firstly, the first break
frequency should be under the UV turnover (Fig 1), especially, for
knots D, A, and B; secondly, a steeper X-ray spectral index than
the optical is needed, so there may be a second break frequency
between UV data and X-ray data; finally, the best fitting model
should explain the flux and index of the X-ray data as well as the
ones of the radio, optical and UV data.

In $\S$ 2, we describe in detail our modified synchrotron model for
knots in the M87 jet. In $\S$ 3, we present and discuss the fitting
results of this model to the wide-band spectra of the knots D, E, F,
A, B, and C1. A summary is given in $\S$ 4.

\section{The Model}

We suggest a synchrotron model which is a modified CI model of K62,
to explain the observed SEDs of the knots in the M87 jet. The
kinetic equation of the relativistic electrons in the CI model (K62)
is $\colon$

\begin{equation}
\frac{\partial N}{\partial t}=\beta\frac{\partial}{\partial
E}(E^2N)+qE^{-p}, \end{equation}

\noindent where $\beta=bB^2_{\perp}$, $b$ is a constant,
$B_{\perp}$ represents the component of the magnetic field
perpendicular to the velocity of the particle. The synchrotron
energy losses are $dE/dt=-\beta E^{2}$. The injection of a
constant spectrum $qE^{-p}$ in the CI model of K62 actually
implies that the acceleration region is the same as the main
emission region in the knots. However, this assumption may not be
true in general, especially for shocks acceleration mechanism
which may be the dominant mechanism of the particle acceleration
in jets. Based on the observational evidence of shocks in the M87
jet (Perlman et al. 1999; PW05; Harris \& Krawczynski 2006), a
more physical scenario would be that the acceleration region and
the main emission region are not strictly co-spatial in the M87
jet, and the accelerated injection electrons may be advected
downstream (this process is similar to the CI model of HM87). The
electrons advection throughout the jet and diffusion throughout
the jet's cross-section with the decrease of the synchrotron
lifetimes from low energies to high energies may result in
spatially stratified emission regions along the jet (M02) and the
consistent narrowing of the jet from radio to optical to X-ray
(PW05). For the acceleration region, a more complex but physical
scenario may be that there are many acceleration sources in a
compact acceleration region which is unresolved by the telescope
beam and each source has a power law distribution of injection
population of relativistic electrons. We assume that all the
sources accelerate electrons under the similar conditions (e.g.,
similar magnetic fields etc.) with the same mechanism which
results in a same spectral index $p$, similar electron number
density, the same maximum energy and fluid velocity for the
injection populations of relativistic electrons into the main
emission region. Before injecting into the main emission region,
the populations of relativistic electrons in the acceleration
region will experience a brief synchrotron loss process. But the
observed knot emission is dominated by the emission from the main
emission region. If the injection electrons into the main emission
region come from a very compact region, not resolved by the
observation, we need to integrate all the contributions of
populations from different locations of the acceleration region by
considering synchrotron losses. We assume that r is the distance
to the main emission region from a source of the acceleration
region. At low energies, the populations from different locations
of the acceleration region have the same spectral shape, so the
integrated injection spectrum of the electrons over the
acceleration region still scales as $E^{-p}$. At higher energies,
the electrons are advected a distance r to the main emission
region before losing a significant fraction of their energy to
synchrotron emission. As the loss time-scale scales as $E^{-1}$,
this distance scales as $E^{-1}$. Thus the integrated injection
spectrum at high energies would be proportional to $E^{-(p+1)}$.
There is a break energy $E_{b}$ between the low energies and the
high energies in the integrated injection spectrum. The integrated
injection spectrum exhibits a cut-off energy $E_{c}$. By adding up
the contributions from sources in the acceleration region, the
integrated injection spectrum as the source into the main emission
region from the acceleration region can be written as,
\begin{equation}
\left\{
\begin{array}{ll}
q_{1}E^{-p},&\mbox{~$E<E_{b}$~;}\\
q_{2}E^{-(p+1)},&\mbox{~$E_{c}>E>E_{b}$~;}\\
0,&\mbox{~$E>E_{c}$~,}
\end{array}
\right. \end{equation}

\noindent where, $q_{2}E^{-(p+1)}$ corresponds to the component of
the injection spectrum at high energies. The spectrum near the
break energy in fact is not a strict power law, and the equivalent
spectral index should be intermediate. But here we would simply
consider a power law form for the spectrum. The real distribution
of the electrons and the magnetic field in the M87 jet is very
complex (M02; Perlman et al. 1999; Perlman et al. 2001b), but we
simply assume that the distribution of the electrons is isotropic
and the large-scale equipartition magnetic field in the main
emission region is approximately symmetrical. The kinetic equation
is$\colon$

\begin{equation}
\frac{\partial N}{\partial t}=\beta\frac{\partial}{\partial
E}(E^2N)+\left\{
\begin{array}{ll}
q_{1}E^{-p},&\mbox{~$E<E_{b}$~;}\\
q_{2}E^{-(p+1)},&\mbox{~$E>E_{b}$~.}
\end{array}
\right. \end{equation}

\noindent Then, we have

\begin{equation}
N(E,\theta,t)=\left\{
\begin{array}{ll}
\frac{q_{1}E^{-(p+1)}}{\beta (p-1)}[1-(1-\beta Et)^{p-1}],&\mbox{~$E<\frac{1}{\beta t}$~;}\\
\frac{q_{1}E^{-(p+1)}}{\beta (p-1)},&\mbox{~$E_{b}>E>\frac{1}{\beta t}$~;}\\
\frac{q_{2}E^{-(p+2)}}{\beta p},&\mbox{~$E>E_{b}$~.}
\end{array}
\right. \end{equation}

For isotropic distributions, we can derive the flux expression of
the modified synchrotron model

\begin{equation}
I_{\nu}\propto\left\{
\begin{array}{ll}
\nu^{-(p-1)/2},&\mbox{~$\nu\ll \nu_{B1}$~;}\\
\nu^{-p/2},&\mbox{~$\nu_{B2}>\nu\gg \nu_{B1}$~;}\\
\nu^{-(p+1)/2},&\mbox{~$\nu>\nu_{B2}$~.}
\end{array}
\right. \end{equation}

$$\nu_{B1}=c_3t^{-2},  \nu_{B2}=c_3
\beta^2 E^2_{b},  c_{3}=3.4\times 10^8 B^{-3},$$

\noindent where $t$ (in yr) is synchrotron lifetime, $\nu_{B1}$ and
$\nu_{B2}$ (in Hz) are the first and second break frequencies,
respectively.

\section{Fitting Results and Discussion}
Now, we apply the above modified synchrotron model to the observed
knot emission in the M87 jet. These include knots D, E, F, A, B, and
C1 (as showed in Figure 1 of Harris \& Krawczynski 2006).

The data we used are listed in Table 1. These include the
published radio to UV data (P01), UV data at $1.8\times10^{15}$ Hz
(WZ05) and X-ray data (PW05). There is another reported X-ray
measurement (M02) made 12 days before PW05 observation. M02 did
not use the same regions to integrate the fluxes for various
components as did P01. PW05, however, did use the same regions as
P01, so we use PW05 data points in our model fitting. There was no
X-ray measurement for the knot C1. So we estimated its flux
density using the measurements for the knot C and the ratio of
knots C1 to C from the optical observation at $1.0\times10^{15}$
Hz by assuming this ratio is the same at both optical
($1.0\times10^{15}$ Hz) and X-ray. There were no error bars in the
X-ray data by PW05. The error bars for the PW05 X-ray data listed
in Table 1 were estimated by assuming the same relative precision
(the uncertainty in terms of a fraction of the value of the
result) in both PW05 and M02 data. We use the weighted least
square method to fit our model (Equation 5) to the wide-band data
in each knot, in this progress a power law form is chosen near the
break frequencies because we only need to concern the trend of the
break frequencies. There are two peak frequencies in our model,
but we don't know where they are when we fit our model to the
broadband data in each knot. So we first arbitrarily divide
broadband data into three groups to perform the least square
method, we could calculate the corresponding reduce chi square
($\chi^2_{\nu}$) by changing the division. All the possible
combinations are considered before we obtain the best fit with a
minimal $\chi^2_{\nu}$ among them. These best-fit parameters for
each knot, are listed in Table 2. The SEDs for the knots in the
M87 jet and the best fits for our model are plotted in Fig 1.

Our model can well fit the radio, optical, UV, and X-ray data for
these knots with three segments except knot E which has two segments
implying that the second break frequency of knot E has exceeded the
X-ray band. It satisfies the aforementioned three constraints for
the SED of the knots of M87 jet (as shown in Fig 1 and Table 2). The
predicted spectra shape of knots from UV to X-ray (from $-p/2$ to
$-(p+1)/2$) could be tested by future telescope in the band. The
equivalent spectral index of the high energy electrons in the
acceleration region may not decay from $p$ to $p+1$, but is likely
to be between them (i.e. the high energy electrons may be near the
break energy). This results in an equivalent spectral index of the
photons possibly between $p/2$ and $(p+1)/2$ (e.g. knot D). Our
model is consistent with the spectral indexes of the knots E, F, A,
B, and C1 within their error bars.

The first break synchrotron frequencies ($\nu_{B1}$) of all the
knots (D, E, F, A, B, and C1) in the M87 jet are under $10^{15}$ Hz
(see Table 2). According to the unified scheme of AGNs and equation
(1) in Bai $\&$ Lee (2001), the Compton components from these knots
will peak at $0.01\sim1$ GeV, and their flux densities at TeV would
be undetectable (based on the synchrotron self-Compton model and the
equation (4) in Bai $\&$ Lee 2001). This implies that the knots D,
E, F, A, B, and C1 are unlikely to be the candidate for TeV emission
in the M87 detected by the HESS observations (Aharonian et al. 2003
$\&$ 2006). The possible candidate for the TeV emission in M87 is
related to the innnermost region of M87 like HST-1 or core itself
(Stawarz et al. 2006; Aharonian et al. 2006; Cheung et al. 2007).
HST-1, the knot closest to the core of M87, has been shown to have a
very dynamic light curve and flaring (Perlman et al. 2003; Harris et
al. 2003; Stawarz et al. 2006; Harris et al. 2006). Its mechanism
may be related to the core and thus more complicated.

From Table 2, as a whole the second break frequency $\nu_{B2}$
decreases along the jet. This may imply that the synchrotron loss in
the acceleration region will increase along the jet.

We find that the value of the particle spectral index $p$ is about
$2.36$ on average, which agrees well with the latest expectations
from both diffusive shock acceleration theory (2.0-2.5, Kirk \&
Dendy 2001) and acceleration by relativistic shocks (2.23 in the
ultrarelativistic limit, Kirk 2002). Particle acceleration at shocks
(e.g., Blandford \& Ostriker 1978) through the first-order Fermi
process is generally believed to occur in jets.

PW05 proposed a possible phenomenological model to modify the CI
model. They suggest that the filling factor of the volume  within
which particles are accelerated declines with increasing energy at
X-ray energy along the jet (not in the direction perpendicular to
the jet), but they can't explain the running mechanism of filling
factor. To explain the curved X-ray spectra of BL Lac objects,
Perlman et al. (2005) consider an episodic particle acceleration
model which assumes only a time-variable particle acceleration.
This results in a logarithmic curvature rather than a sudden break
and could be related to the broadband spectral shape too.
Fleishman (2006) suggests a very different model which explicitly
takes into account the effect of the small-scale random magnetic
field, probably present in the M87 jet. But the energy densities
contained in small-scale and large-scale magnetic fields may be
incomparable, we think that the electrons in the large-scale
magnetic field could also give rise to emission of the knots in
our model, especially at high frequencies. The idea of a secondary
population of the relativistic electrons that have different
spectral index from the first population is discussed by Jester et
al. (2005), which is partially similar to our model. But we
consider a lot of populations of relativistic electrons that have
the same spectral index in the acceleration region and discuss the
detailed process that may be responsible for the knots in the M87
jet.

\section{Conclusion}

We propose a modified CI model that considers a decay of spectral
index of injection electrons possibly due to the sum of the
injection spectrum from different acceleration sources with
synchrotron losses in the thin acceleration region, so there are
two break frequencies at two sides of which the spectral index
changes for the spectra of knots in the M87 jet. We consider that
the emission of the knot may be still emitted by the relativistic
electrons in the large-scale magnetic field at high frequencies as
well as the low frequencies, but not by the small-scale random
magnetic field (e.g., Fleishmann 2005). Our model gives a
satisfactory fit to the SEDs of knots in the M87 jet. Based on our
analysis, the knots D, E, F, A, B, and C1 are unlikely to be
responsible for the TeV emission detected in M87. The fitting
results from our model imply that the particles in M87 are
accelerated by shocks, and as a whole the second break frequencies
of knots decrease down the jet. We also predict the spectra of
knots from UV to X-ray, which could be tested by future
observations in the band.

We thank C. Z. Waters for helpful communications and supplying his
code to us for reference. Many thanks are due to the referee for
the critical and constructive comments, which improved our work
greatly. This work was supported in part by the National Natural
Science Foundation of China (grants 10573029, 10625314, and
10633010) and the Knowledge Innovation Program of the Chinese
Academy of Sciences (Grant No. KJCX2-YW-T03), and sponsored by the
Program of Shanghai Subject Chief Scientist (06XD14024). Z.-Q.
Shen acknowledges the support by the One-Hundred-Talent Program of
the Chinese Academy of Sciences.

\clearpage

\begin{landscape}
\setlength{\headsep}{1cm}
\begin{deluxetable}{crrrrrrrr}
\tablecolumns{7} \tabletypesize{\footnotesize} \tablecaption{Flux
Densities of Knots in the M87 Jet\label{tbl-1}} \tablewidth{0pt}
\tablehead{ \multicolumn{1}{c}{} &
\multicolumn{6}{c}{Flux Density}\\
\cline{2-7}\\
\colhead{Frequency} & \colhead{D} & \colhead{E} & \colhead{F} &
\colhead{A} & \colhead{B} & \colhead{C1}& \colhead{Observing} & \colhead{Reference}\\
\colhead{$\nu$ (Hz)} & \multicolumn{6}{c}{} & \colhead{Date}}
\startdata
\colhead{{\it VLA} (mJy):}\\
$1.50\times10^{10}$ 
& $161.54\pm1.92$ & $48.05\pm0.81$ & $144.90\pm1.86$ & $1218.00\pm12.00$ & $808.40\pm8.30$ & $544.70\pm5.60$ & 94 Feb 04& P01 \\
\colhead{{\it HST} ($\mu$Jy):}\\
$1.45\times10^{14}$ 
& $280\pm11$ & $71.4\pm3.0$ & $262\pm11$ & $2633\pm105$ & $1739\pm70$ & $971\pm39$& 98 Apr 04& P01 \\
$1.87\times10^{14}$ 
& $224\pm9$ & $67.2\pm2.0$ & $271\pm11$ & $2344\pm90$ & $1489\pm60$ & $797\pm32$ & 98 Feb 26& P01\\
$2.66\times10^{14}$ 
& $168\pm7$ & $42.9\pm2.3$ & $147\pm6$ & $1829\pm73$ & $1104\pm44$ & $606\pm24$ & 98 Feb 26& P01\\
$3.75\times10^{14}$ 
& $150\pm4$ & $42.3\pm1.3$ & $158\pm5$ & $1363\pm34$ & $811\pm20$ & $415\pm10$ & 98 Feb 25& P01\\
$4.50\times10^{14}$ 
& $117\pm3$ & $33.4\pm0.9$ & $123\pm3$ & $1086\pm27$ & $623\pm16$ & $310.7\pm7.8$ & 98 Feb 25& P01\\
$6.58\times10^{14}$ 
& $96.2\pm2.5$ & $28.0\pm0.8$ & $98.8\pm2.6$ & $904\pm23$ & $504\pm13$ & $241.0\pm6.1$ & 98 Feb 25& P01\\
$1.00\times10^{15}$ 
& $59.5\pm1.6$ & $16.2\pm0.6$ & $62.7\pm1.7$ & $586\pm15$ & $306.8\pm7.8$ & $135.9\pm3.5$ & 98 Feb 25& P01\\
$1.80\times10^{15}$ 
& $26.7\pm0.9$ & $10.0\pm0.9$ & $34.5\pm1.7$ & $275.2\pm12.7$ & $147.4\pm7.9$ & $80.8\pm12.9$ & 01 Feb 23& WZ05 \\
\colhead{{\it Chandra} (nJy):}\\
$2.42\times10^{17}$ 
& $51.5\pm4.2$ & $32.2\pm6.5$ & $20.1\pm5.2$ & $156\pm8.8$ & $30.3\pm5.5$ & $14.6\tablenotemark{a}\pm5.2$& 00 Jul 29& PW05\tablenotemark{b}\\
\enddata
\tablenotetext{a}{The X-ray flux density of knot C1 was estimated
by assuming the same flux ratio of the knot C1 to the knot C at
both optical ($1.00\times10^{15}$) Hz and X-ray band.}
\tablenotetext{b}{The error bars for the PW05 X-ray data were
estimated by assuming the same relative precision in both PW05 and
M02 data (see text).}
\end{deluxetable}
\end{landscape}

\clearpage
\begin{figure}
\includegraphics[scale=1.5]{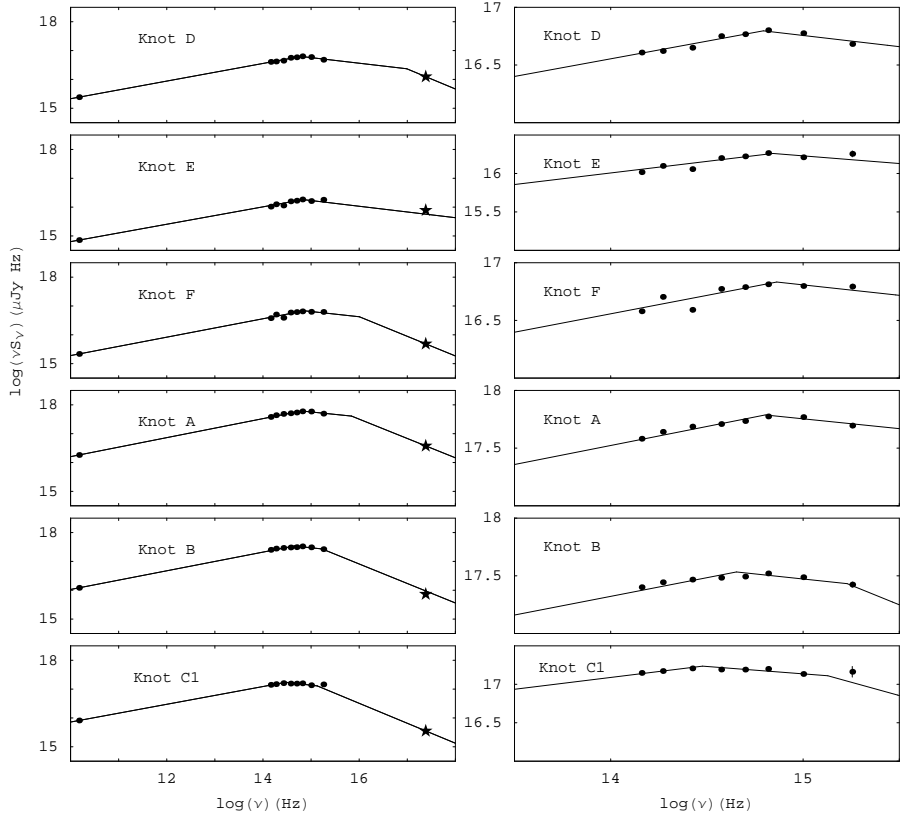}
\caption{Left panel: plot of the SEDs for the knots in the M87 jet
from radio through X-ray wavelengths. Right panel: plot zooming in
on the optical and UV region, where most of the data points are
and also where most of the curvature is. The data from radio to UV
are plotted as filled circles, the X-ray data from PW05 as filled
star. The solid lines display model fits that include the X-ray
data from PW05. The error bars of most measurements are too small
to be seen here. The best-fit parameters are listed in Table 2.}
\end{figure}

\begin{deluxetable}{lccccc}
\tabletypesize{\small} \tablewidth{0pc} \tablecaption{Parameters for
Model Fits for Radio through X-ray Data \label{tab:modelX}}
\tablehead{\colhead{Parameter}  & \colhead{$\nu_{B1}$($10^{14}$Hz)}
& \colhead{$\nu_{B2}$($10^{15}$Hz)} &
  \colhead{$p$} & \colhead{$\chi^2_{\nu}$}}
  \startdata
  knot D & $6.21$ & $97.5$ & $2.39$ & $2.34$\\
  knot E & $6.97$ & $\cdots$ & $2.40$ & $3.73$ \\
  knot F & $7.30$ & $10.1$ & $2.36$ & $9.33$\\
  knot A & $6.35$ & $6.94$ & $2.34$ & $1.89$\\
  knot B & $4.50$ & $1.69$ & $2.35$ & $4.39$\\
  knot C1 & $3.00$ & $1.34$ & $2.38$ & $2.55$\\
\enddata
\tablecomments{Col. (1): Knot designation. Col. (2): First break
frequency in $10^{14}$ Hz. Col. (3): Second break frequency in
$10^{15}$ Hz. Col. (4): Spectral index of electrons. Col. (5):
Reduced chi square.}
\end{deluxetable}

\end{document}